\documentclass[aps]{revtex4}
\usepackage{amsfonts}
\usepackage{graphicx}
\usepackage{pifont}
\usepackage{graphicx}
\usepackage{amsmath,amssymb,amsthm}

\newcommand{\ini}{\begin{equation}}
\newcommand{\fin}{\end{equation}}
\newcommand{\inia}{\begin{eqnarray}}
\newcommand{\fina}{\end{eqnarray}}

\begin{document}

\title{\textbf{Localizing Gauge Fields on a Topological Abelian String and
the Coulomb Law.}}
\author{Rafael S. Torrealba\ S.}
\thanks{rtorre@ucla.edu.ve}
\affiliation{Departamento de F\'{\i}sica. Universidad Centro Occidental "Lisandro
Alvarado"}

\begin{abstract}
The confinement of electromagnetic field is studied in axial symmetrical,
warped, 6D World Brane, using a recently proposed topological abelian string
vortex solution as background. It was found, that the massless gauge field
fluctuations follow 4D Maxwell equations in the Lorenz gauge. The massless
zero mode is localized when the thickness of the string-vortex is less than 5%
$\beta $/4$\pi e^{2}v^{2}$ there are not others localized massless modes.
There is also an infinite of non localized massive Fourier modes, that
follow 4 dimensional Proca equations with a continuous spectrum. To compute
the corrections to the Coulomb potential, a radial cutoff was introduced, in
order to achieve a discrete mass spectrum. As main result, a $\frac{R_{o}}{%
\beta R^{2}}$ correction was found for the 4D effective Coulomb law, the
result is in correspondence with the observed behavior of the Coulomb
potential at nowadays measurable distances.

\vspace{0.5 cm} PACS numbers: 04.20.-q, 11.27.+d, 04.50.+h
\end{abstract}

\maketitle



\section{Introduction}

In the brane world view, the universe is considered a 3-brane contained in
some topological defect. This 3-brane is embedded in non compact, large,
extra dimensional spacetime with finite volume \cite{ADD} usually with
warping metric \cite{RS1} in the directions of the additional dimensions.
Most of the research had been done \cite{ReferencesDW}\ on domain walls in 5
dimensions, due to the fact that this is the simplest kind of topological
defect and exact solutions are available for both the metric and scalar
fields. Important advances had been made on domain wall's brane world, as
possible explanation to the hierarchy problem \cite{RS1}, localization of
gravity \cite{RS2}, and the confinement of some fundamental fields \cite%
{Bajc} as chiral fermions \cite{Rubakov-Shaposnikov}, but there are still
several unsolved problems as the localization of gauge fields \cite%
{DvaliShifmann}\cite{Bajc}, stability under casimir energy or quantum
fluctuations \cite{StabilityDW} and the lack of a complete\ supersymmetric
version.

Its is widely known that in 5 dimensional brane worlds, the gauge and spinor
fields can scape from the 3 brane universe into the bulk. Yukawa coupling
gives a natural way of localize fermionic fields on topological defects \cite%
{Rubakov-Shaposnikov} but the confinement of spin 1 fields remains as an
open problem. Some mechanisms as DS \cite{DvaliShifmann} and DGS \cite%
{DvaliGabadadzeShifmann} of quasi localization has been proposed. These
mechanism seem to unnaturally force, or increase the effect of the "wall"
over the volumetric "bulk" terms. To do that, they introduce two different
coupling constant for the electromagnetic field and very strict restrictions
on the fields and parameters. Some recent calculations indicate very strict
bounds on the wall thickness of many orders of magnitude thinner than the
Planck scale \cite{ThickDGS} to achieve a 4 dimensional behavior. The loss
of electromagnetic radiation at low frequency and the non preservation of
charge are also drawbacks related to the confinement of gauge fields in
domain wall brane worlds. Although this drawbacks can be controlled in
several ways, they are also unwanted features for a theory modeling the
universe.

Some years ago, brane worlds had been proposed over more complicated
topological defect than domain walls. For example world brane had been
proposed for the 6 dimensional vortex \cite{Shaposhnikov}\cite{RSVortex} and
7 dimensional monopole \cite{RSmonopole}\cite{Hedgehog}. The confinement of
spin 1 gauge fields \cite{IchiroOda} seems to be possible to achieve in
string-vortices in 6 dimensions: it has been found that electromagnetic
gauge field fluctuations has a localized zero mode \cite{IchiroOda}\cite%
{Giovannini1}, however the graviphoton zero modes (spin 1 fluctuations of
the metric) are not normalizable due to its behavior either at infinity or
very near to the string-vortex \cite{Giovannini2}\cite{FluctuationBW}.\
These works are based on numerical or asymptotical approximations, because
there are not known general analytic solutions for curved 6D string-vortices
and very few exact solutions had been reported \cite{belga}\cite{GRG2010}.
Moreover, almost nothing is known about the non zero modes and its influence
on the propagators, potentials and interactions.

Recently, new Randall Sundrum scenarios in 6 dimensional curved space time
based on exact topological solutions to the abelian higgs model \cite%
{GRG2010} had been reported. These solutions correspond to different
Einstein Maxwell field vacua, with scalar kink solitons and auto-interaction
potential, that are in fact, electromagnetic uncharged string-vortices with
non trivial winding number. These topological abelian strings exhibits a
localized spin 2 zero mode, that gives rise to the $\frac{1}{R}$ newtonian
gravity, while the rest of the modes gives account of a $\frac{1}{R^{3}}$
correction to the potential \cite{Shaposhnikov}. Is the purpose of these
paper to study the confinement of spin 1 gauge field, on a background of a 6
dimensional topological abelian string vortex, in order to establish the
localization or not, of all the photon's fourier modes and calculate the
corrections to the Coulomb law.

The organization of the paper is as follows: section II explains the model:
the topological abelian string vortex. Section III reviews the confinement
of the zero mode of linearized gravity and the non confinement of the other
modes. In section IV, the gauge fixing conditions and the decoupling from
the graviphotons and graviscalars field are discussed and the 4D Maxwell
equations recovered in a perturbative approach using the topological abelian
string as background. In section V, the gauge field equations are expanded
in Fourier modes. A localized, normalizable, solution to the massless zero
mode is obtained numerically in the thin string approximation: $\delta <%
\frac{5\beta ^{2}}{4\pi e^{2}v^{2}}$ it is shown that there are not other
localized massless modes. In section VI the massive modes are studied, a
regularization is proposed based on the introduction of a radial cutoff and
the corrections to the Coulomb law calculated. It is the main result of this
paper a $\frac{R_{o}}{\beta R^{3}}$ correction to the Coulomb $\frac{1}{R}$
potential. In section VII the range of validity of this corrections and its
adjustment to known experimental bounds is discussed, finally some
conclusions and\ remarks are presented.

\section{A Topological Einstein Abelian String-Vortex Solution.}

We start from the 6 dimensional action%
\begin{equation}
S=-\int dx^{6}\sqrt{-G}\left[ \frac{{\ 1}}{{\ 2\chi }}R+\Lambda \right]
+\int dx^{6}\sqrt{-G}\left[ \frac{{\ 1}}{{\ 2}}(D^{A}\phi )^{\ast
}(D_{A}\phi )-\frac{{\ 1}}{{\ 4}}F_{AB}F^{AB}-V\right] ,  \label{EinsteinAH}
\end{equation}%
where the first integral is the 6D Einstein Hilbert action, with a bulk
cosmological constant $\Lambda $ and $D_{B}=\partial _{B}-ieA_{B}$ is the
covariant derivative.

In what follows we will set $\ \chi =\frac{8\pi }{(M_{6})^{4}}=1,$ where $%
M_{6}$ \ is the 6D Planck mass and we will follow the notation in \cite%
{RSVortex}\footnote[1]{$A,B$ and uppercase Latin indices run over 4+2 dim. $%
\mu ,\nu $ and Greek indices over 4 dim. middle alphabet Latin indices run
in 3 dim $i,j=0,r,\theta $ and firsts alphabet Latin indices run over 2 dim $%
a,b=r,\theta .$ Bold $\boldsymbol{\mu }$ was used for mass}.

Here we will look for a geometry 2+3+1 composed by a 3-brane that contains
the 3+1 physical universe and 2 extra dimensions, where we can choose
coordinates $(r,\theta )$ and the metric is given by:%
\begin{equation}
ds^{2}=M^{2}(r)\ \eta _{\mu \nu }dx^{\mu }dx^{\nu }-R_{o}^{2}\
L^{2}(r)d\theta ^{2}-dr^{2},  \label{metric}
\end{equation}%
with $\eta _{\mu \nu }=(+,-,-,-)$. In this context $r\in \lbrack 0,\infty )$
and $\theta \in \lbrack 0,2\pi ]$ are the coordinates of the extra
dimension, $R_{o}^{2}L(r)$ acts as a radial factor and $M(r)$ is a warp
factor, rapidly diminishing when moving away from a 4 dimensional 3-brane
located at $r=0$.

Assuming that the scalar and gauge fields in (\ref{EinsteinAH}) depends on
the extra coordinates as in the Nielsen Olesen ansatz \cite{Nielsen-Olesen}%
\cite{Shaposhnikov}

\begin{eqnarray}
\phi (x^{\mu },r,\theta ) &=&v\ f(r)\ e^{in\theta },\text{ \ \ \ \ \ \ \ \ \
\ }\ \ n\in \mathbb{Z}  \label{F} \\
A_{\theta }(x^{\mu },r,\theta ) &=&\ \frac{n-P(r)}{e},\text{\ \ \ \ }
\label{Atheta} \\
A_{r}(x^{\mu },r,\theta ) &=&0.  \label{Ar} \\
A_{\mu }(x^{\mu },r,\theta ) &=&0\text{ \ \ }  \label{Amu}
\end{eqnarray}%
where the $v=1$ is dimensional factor with units $lenght^{-2}$. Performing
variations in the action (\ref{EinsteinAH}) we get the curved version of
Nielsen Olesen vortex equations \cite{RSVortex}:
\begin{eqnarray}
\frac{d^{2}f}{dr^{2}}+(4m+l)\frac{df}{dr}-\frac{P^{2}\ f}{R_{o}^{2}\ L^{2}}-%
\frac{1}{v^{2}}\frac{dV}{df} &=&0,  \label{CurvedVotexA} \\
\frac{d^{2}P}{dr^{2}}+(4m-l)\frac{dP}{dr}-v^{2}e^{2}f^{2}P &=&0,
\label{CurvedVortexB}
\end{eqnarray}%
where
\begin{equation*}
P=n\ [1-\alpha (r)];\text{\ \ \ \ \ \ \ \ \ }m=\frac{d}{dr}\ln [M(r)];\text{
\ \ \ \ \ \ \ \ }l=\frac{d}{dr}\ln [L(r)].
\end{equation*}

String-vortex solutions are classically obtained with the boundary
conditions:

\begin{equation}
\begin{array}{cc}
f(r\rightarrow 0)=0, & f(r\rightarrow \infty )=1, \\
\alpha (r\rightarrow 0)=0, & \alpha (r\rightarrow \infty )=1,%
\end{array}
\label{Boundary}
\end{equation}%
by numerics and asymptotical methods. Very few solutions to the equations (%
\ref{CurvedVotexA},\ref{CurvedVortexB}) are known. Indeed only the
Bogomoln'yi solution for flat space in the critical case \cite{Bogomoln'yi}
\cite{Vega} is known to be exact. Recently in \cite{belga}\cite{GRG2010} the
boundary condition $\alpha (r)=1$ was used to find new exact solutions, so
we will assume here:%
\begin{equation}
P(r)\equiv 0,\text{ \ }\ \ \forall \text{\ \ \ }\ r\in (0,\infty )  \label{P}
\end{equation}%
and the equation (\ref{CurvedVortexB}) is automatically satisfied.

Boundary condition (\ref{P}) lead to a apparent singularity at in the
vectorial field because the direction $\hat{u}_{\theta }$ is not defined at
r=0, and the potential 1-form is locally defined by%
\begin{equation}
A=A_{B}dx^{B}=A_{\theta }d\theta .  \label{Atita}
\end{equation}%
The model (\ref{EinsteinAH}) is invariant under the group U(1) of local
gauge transformation:%
\begin{equation*}
\phi \rightarrow e^{i\vartheta (x)}\phi \qquad \qquad A_{B}=A_{B}+\frac{1}{e}%
\partial _{B}\vartheta (x)
\end{equation*}%
So, when the phase of the scalar field is choose as
\begin{equation*}
\vartheta (x,r,\theta )=n\ \theta \qquad \Longrightarrow \qquad \delta
A_{\theta }=\ \frac{n}{e}
\end{equation*}%
that is exactly (\ref{Atheta}) when $\alpha (r)=1$ and is valid for $r\neq 0$%
.

Equation (\ref{Atita}) generates a null electromagnetic field $F_{AB}=0,$
and although this type of string-vortex has neither electric nor magnetic
charge, it is a topological vortex solution because it still has a non
trivial integer winding number:

\begin{equation}
n=\frac{e}{2\pi }\int\limits_{C}A,\qquad n\in
\mathbb{Z}
\label{homotopy1}
\end{equation}%
where $C$ is any closed curve around a "string" at $r=0$.

It is not possible to continuous pass or continuously deforms a curve with a
particular value $n$ to a curve with a different $n$. For each homology
class of curves $C$ we will get the same integer value for $n$. Each
different value $n$ will correspond to a class of curves $C$ and labels an
specific topological vacuum.

This kind of string-vortex solutions will be referred as "topological
string-vortices" \cite{GRG2010} and they are useful to construct 6
dimensional brane world. In order to obtain a feasible Randall Sundrum
scenario in 6 D, we still have to prove that that the metric factor $%
M^{2}(r) $ is warped and the potential shows breaking of the symmetry.

Einstein equations are obtained performing variations of the metric in the
action and \ jointly with (\ref{CurvedVotexA}) conforms a system of 4
coupled non linear differential equations and 4 variables $(m,l,f,V).$ With
straightforward combinations, the complete system could be written as:%
\begin{eqnarray}
f^{\prime \prime }+(4m+l)f^{\prime }-\frac{1}{v^{2}}\frac{d}{df}V &=&0,
\label{system1} \\
l^{\prime }+(4m+l)l &=&-\frac{\chi }{2}V,  \label{system2} \\
m^{\prime }+(4m+l)m &=&-\frac{\chi }{2}V,  \label{system3} \\
m^{\prime }+m^{2}-ml &=&-\frac{1}{4}(f^{\prime })^{2}.  \label{system4}
\end{eqnarray}%
Where the 6D (bulk) cosmological constant was absorbed into the redefinition
of the potential $V\longrightarrow V+\Lambda $ and although $\chi =1$ and $%
v=1$, had been introduced in the equations to match the dimension of the
system of units.

Equations (\ref{system2}) and (\ref{system3}) implies $m=l$, so the system
reduces to:
\begin{eqnarray}
f^{\prime \prime }+5m.f^{\prime }-\frac{1}{v^{2}}\frac{d}{df}V &=&0,
\label{Efe} \\
m^{\prime }+5m^{2} &=&-\frac{\chi }{2}V,  \label{m} \\
m^{\prime } &=&-\frac{1}{4}(f^{\prime })^{2}.  \label{mprima}
\end{eqnarray}%
for a given $V(f)$ the system (\ref{Efe},\ref{m},\ref{mprima}) is not longer
independent, and always one of the equations can be obtained from the other
two. But solving this system is not an easy task, due to its coupled non
linear nature.

We will follow here the approach developed in \cite{grg} and \cite{prd},
more details could be found in \cite{GRG2010}. Instead of try to solve for a
given $V(f),$ we will give a probe function $f(r)$ that accomplishes the
boundary conditions.

So, from (\ref{mprima}) it is easy to obtain $m$ as:%
\begin{equation}
m=-\frac{1}{4}\int dr(f^{\prime })^{2}-k_{RS},  \label{integralM}
\end{equation}%
where $k_{RS}$ is an additional Randall Sundrum constant warp factor. Then,
from equation (\ref{m}) the potential $V(r)$ could be obtained as a function
of $r$. In order to obtain the interaction potential $V(f)$ we must solve $%
r(f)$ from the probe function $f(r)$ so:
\begin{equation}
V(f)=V(r(f)),  \label{Vf}
\end{equation}%
Of course a solution to be physically acceptable must have a stable
potential with spontaneous breaking of symmetry, as we will show immediately.

Equations (\ref{Efe},\ref{m},\ref{mprima}) are very similar to that of 5D
domain walls in \cite{grg}, so we will try:
\begin{equation}
f=f_{0}\arctan \left( \sinh \ \beta r/\delta \right) ,\qquad f_{0}=2\sqrt{%
\delta },  \label{ffo}
\end{equation}%
using (\ref{integralM}) we get
\begin{equation}
m=-\beta \tanh \left( \beta r/\delta \right) ,\text{ \ where we set\ \ \ }%
k_{RS}=0,  \label{mchica}
\end{equation}

That corresponds to the metric "warped" factor
\begin{equation}
M(r)=L(r)=\cosh ^{-\delta }(\beta r/\delta ),\qquad \delta >0,\qquad \beta
>0.  \label{Mgrande}
\end{equation}

From equation (\ref{m}) we obtain the potential $V=V(r)$ as:
\begin{equation}
V(r)=\frac{2\beta ^{2}}{\delta }[(1+5\delta ){sech}^{2}(\beta r/\delta
)-5\delta ],  \label{VdeR}
\end{equation}%
and solving $r$ as a function of $f$ by means of (\ref{ffo}) we have $\cos
^{2}(\frac{f}{f_{o}})={sech}^{2}(\frac{\beta r}{\delta })$. So finally the
interaction potential for the scalar field is
\begin{equation}
V(f)=\frac{2\beta ^{2}}{\delta \chi }[(1+5\delta )\cos ^{2}(f/f_{o})-5\delta
].  \label{VdeF}
\end{equation}

\begin{figure}[h]
\begin{minipage}[h]{0.3\linewidth}
\includegraphics[width=4cm,angle=0]{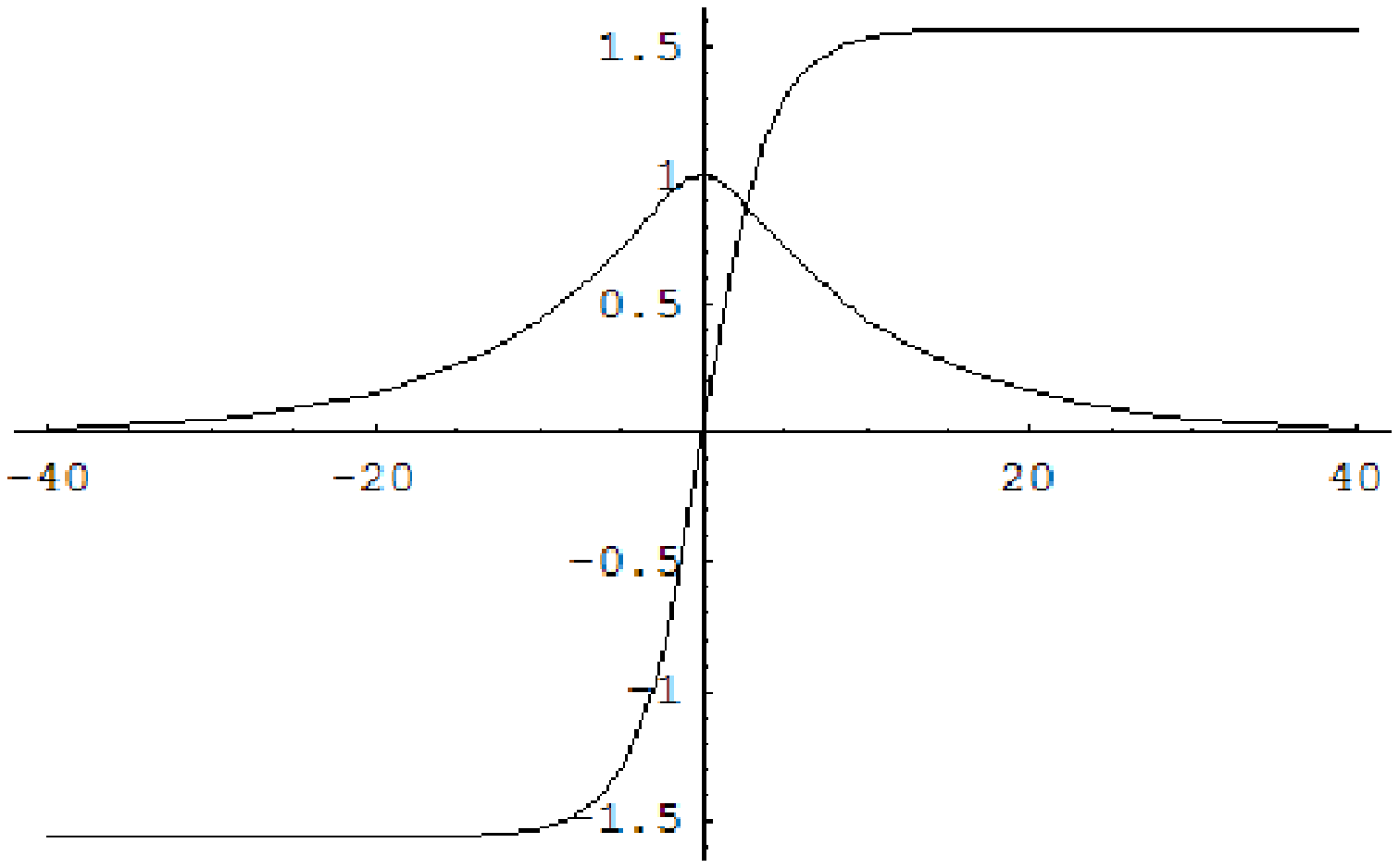}
\end{minipage}%
\begin{minipage}[h]{0.3\linewidth}
\includegraphics[width=4cm,angle=0]{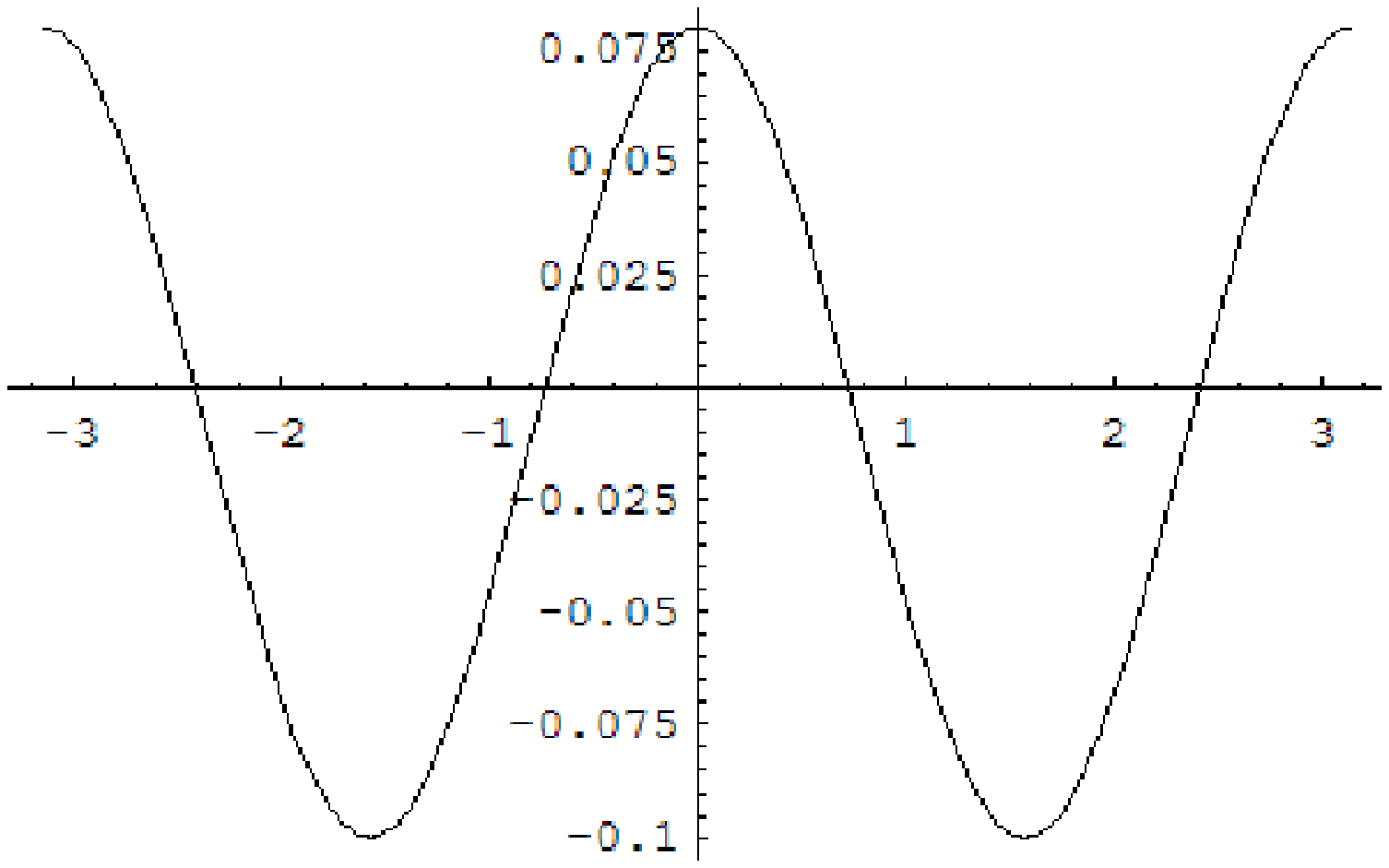}
\end{minipage}%
\begin{minipage}[h]{0.3\linewidth}
\includegraphics[width=4cm,angle=0]{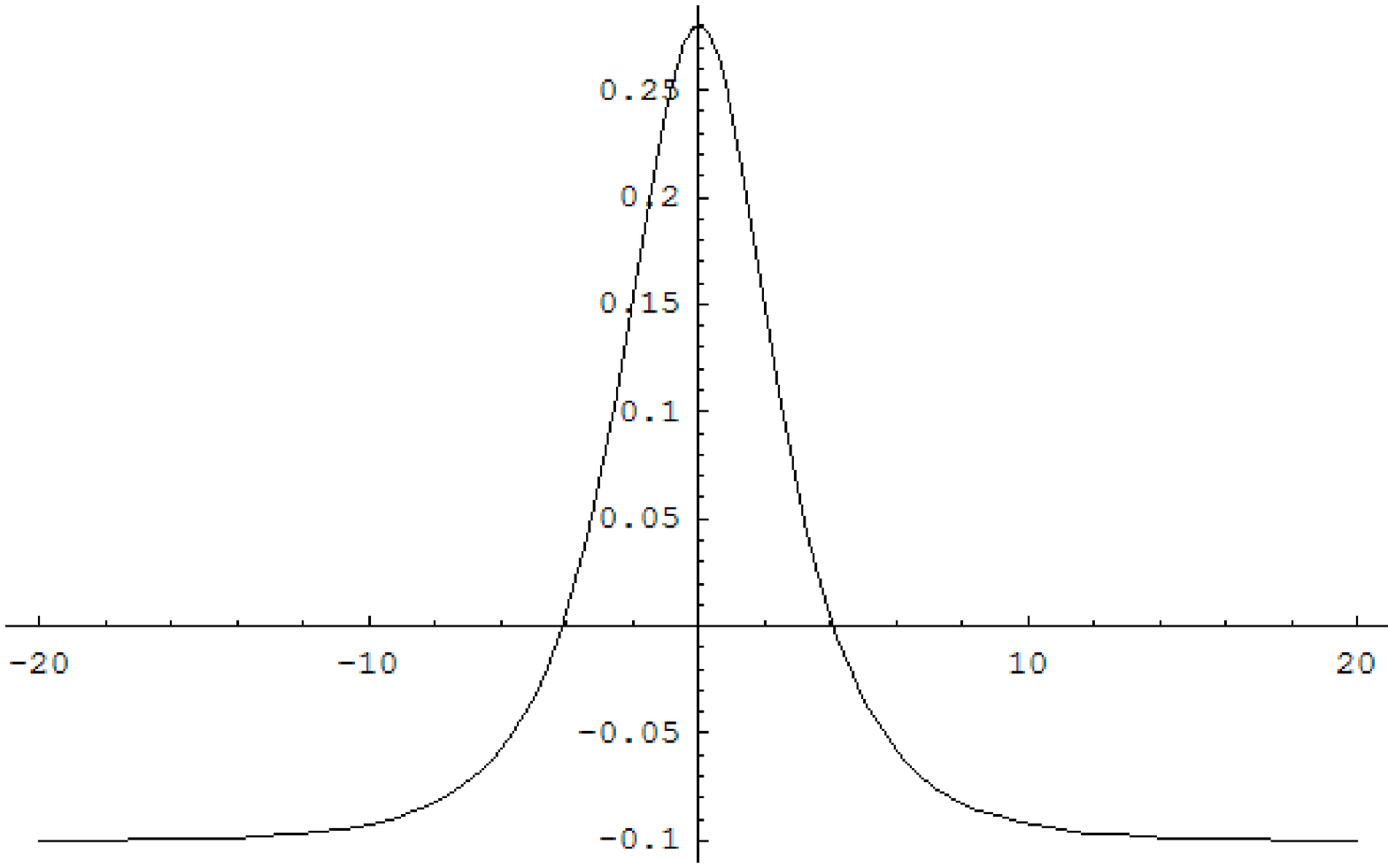}
\end{minipage}
\caption{Graphic of the metric factor M, scalar field (left), and the energy
density (right)vs $r$. Graphic of the auto-interaction potential vs $f$
(center). $fo=1 $, $\protect\delta =1/4,$ $\protect\beta =0.1,\protect\chi %
=1 $}
\label{MetricPotential}
\end{figure}

This is an stable potential, with spontaneous symmetry breaking minima,
where the scalar field interpolates smoothly between $f=\pm \pi \sqrt{\delta
}$ in AdS space-time with cosmological constant $\Lambda =-10\beta ^{2}/\chi
$ as could be seen in Fig.\ref{MetricPotential}.

\section{Localized Gravity on the Topological Abelian String}

We will now study linearized spin 2 metric fluctuations from the metric (\ref%
{Mgrande}) given at first order by

\begin{equation}
g_{\nu \tau }\approx M^{2}[\eta _{\nu \tau }+h_{\nu \tau }]
\end{equation}%
we look for eigenfunctions with separated variables of the form:
\begin{equation}
h_{\tau \nu }(x^{\upsilon },r,\theta )=h_{\tau \nu }(x^{\upsilon
})\sum_{\kappa }\phi _{\boldsymbol{\mu }}(r)e^{i\kappa \theta },
\label{fluctuacionG}
\end{equation}%
Then the gravitation equation from (\ref{EinsteinAH}) splits into two
equations as in \cite{Shaposhnikov}

\begin{equation}
\partial _{\upsilon }\partial ^{\upsilon }h_{\nu \tau }(x^{\upsilon
})=m_{o}^{2}\ h_{\nu \tau }(x^{\upsilon })  \label{LaplaceBeltrami}
\end{equation}%
\begin{eqnarray}
-\frac{1}{M^{2}L}\frac{\partial }{\partial r}\left[ M^{2}L\left( M^{2}\frac{%
\partial }{\partial r}(\phi _{\boldsymbol{\mu }})\right) \right] &=&%
\boldsymbol{\mu }^{2}\phi _{\boldsymbol{\mu }}  \label{wave} \\
\boldsymbol{\mu }^{2} &=&m_{o}^{2}-\frac{\kappa ^{2}}{R_{o}^{2}}
\end{eqnarray}%
where $m_{o}^{2}$ is the constant of the separation of the equation for the
gravity in variables $(x^{\upsilon })$ and $(r,\theta )$. Here the
eigenvalues $\boldsymbol{\mu }$ are effective mass factors that depends on $%
m_{o}$ and $\kappa $.

Looking for solutions of the form of flat waves:%
\begin{equation*}
h_{\tau \nu }(x)=\eta _{\tau \nu }e^{ip.x}
\end{equation*}%
equation (\ref{LaplaceBeltrami}) could be interpreted, as massive graviton
with mass term $p_{\nu }p^{{\nu }}=m_{o}^{2}\geq 0$ in 4 dimension. Equation
(\ref{wave}) could be though, as the radial equation of a massive scalar
mode, in curved 2 dimensional space $(r,\theta )$ with an effective mass
given by $\boldsymbol{\mu }^{2}\geq 0$, that is also non negative because
also $p_{A}p^{A}\geq 0$ in 6 dimensions.

Upon substitution of (\ref{Mgrande}) and using $M(r)=L(r),$ we get the
equation for massive metric fluctuations:%
\begin{equation}
-\phi _{\boldsymbol{\mu }}^{\prime \prime }+5\beta \tanh (\beta r/\delta
)\phi _{\boldsymbol{\mu }}^{\prime }-\boldsymbol{\mu }^{2}\cosh (\beta
r/\delta )\phi _{\boldsymbol{\mu }}=0,  \label{massive}
\end{equation}%
So integrating the equation (\ref{massive}) when $\boldsymbol{\mu }^{2}=0$,
we obtain the massless mode
\begin{equation}
\phi _{o}=k_{o}+k_{1}\int dr\cosh ^{5\delta }(\beta r/\delta ).  \label{Phi0}
\end{equation}%
with integrations constants $\ k_{o}$ \ and\ $\ k_{1}$.

As $\cosh ^{5\delta }[\beta r/\delta ]$ is monotonous growing, we must fix $%
k_{1}=0$, in order to render $\phi _{o}$ \ normalizable. That is consistent
with the boundary conditions%
\begin{equation}
\phi _{\boldsymbol{\mu }}^{\prime }(0)=\phi _{\boldsymbol{\mu }}^{\prime
}(\infty )=0,
\end{equation}%
that allows (\ref{wave}) to be (\ref{massive}) a Sturm Liouville well posed
problem with weight function $M^{2}L=\cosh ^{-3\delta }(\beta r/\delta )$

\begin{figure}[h]
\begin{minipage}[h]{0.50\linewidth}
\includegraphics[width=5cm]{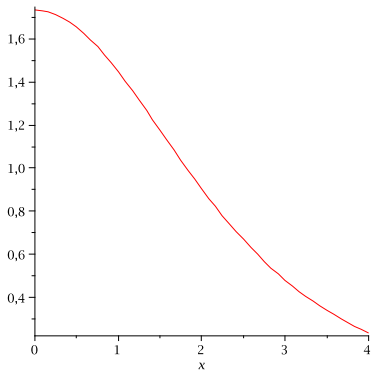}
\end{minipage}
\caption{Graviton massless zero mode Eq.(\protect\ref{FiCero}): $\boldsymbol{%
\protect\mu }=0,$ $\protect\beta =1,\protect\delta =1$ }
\label{Fig2}
\end{figure}

The ortonormalization condition to be satisfied by $\phi _{\boldsymbol{\mu }%
} $ is \cite{Shaposhnikov}%
\begin{equation}
\int_{0}^{\infty }dr\ M^{2}L\ \phi _{\boldsymbol{\mu }}^{\ast }(r)\ \phi _{%
\boldsymbol{\nu }}(r)=\int_{0}^{\infty }dr\cosh ^{-3\delta }(\beta r/\delta
)\ \phi _{\boldsymbol{\mu }}^{\ast }\ \phi _{\boldsymbol{\upsilon }}=\delta
_{\boldsymbol{\mu \upsilon }},  \label{normaG}
\end{equation}%
so the equivalent wavefunction in 1 dimensional quantum mechanics is

\begin{equation}
\psi _{\boldsymbol{\mu }}(r)=\cosh ^{-3\delta /2}(\beta r/\delta )\ \phi _{%
\boldsymbol{\mu }}(r).  \label{FiMiu}
\end{equation}%
Finally, the massless normalized equivalent wavefunction is given by:%
\begin{equation}
\psi _{o}(r)=k_{o}\cosh ^{-3\delta /2}(\beta r/\delta ).  \label{FiCero}
\end{equation}%
This function is strongly decaying, as seen in Fig.\ref{Fig2}. So we
conclude that the massless, spin two, gravitation mode is localized on the
3-brane and strongly concentrated around $r=0$ as we expected for a RS
scenario.

Although general wavefunction solutions are rather cumbersome, what is
really important is the asymptotic behavior of the massive modes. Far from
the vortex core or in the thin domain wall limit, when$\ \delta \rightarrow
0 $, we can approximate (\ref{Mgrande}) by%
\begin{equation}
M=\cosh ^{-\delta }(\beta r/\delta )\cong (\frac{{1}}{{2}})^{\delta }\
e^{-\beta r},  \label{aproximation}
\end{equation}

So the massive wavefunction (\ref{massive}) could be approximated by%
\begin{equation}
\psi _{\boldsymbol{\mu }}(r)\cong e^{-\frac{3}{2}\beta r}\ \phi _{%
\boldsymbol{\mu }}(r),
\end{equation}%
and the localized zero mode by%
\begin{equation}
\psi _{o}(r)\cong \sqrt{3\beta }e^{-\frac{3}{2}\beta r}\ ,
\end{equation}

The massless zero mode is then localized in the vicinity of $r\cong 0$, that
is on the 3 brane where the known universe is located, and decays
exponentially when when $r$ is increased.

Using the approximation (\ref{aproximation}) and (\ref{FiMiu}) into the
differential equation (\ref{massive}), and taking the limit $\beta r/\delta
\rightarrow \infty $ for which $\tanh (\beta r/\delta )\rightarrow 1,$ we
obtain%
\begin{equation}
-\phi _{\boldsymbol{\mu }}^{\prime \prime }+5\beta \ \phi _{\boldsymbol{\mu }%
}^{\prime }-\boldsymbol{\mu }^{2}e^{2\beta r}\phi _{\boldsymbol{\mu }}=0,
\label{PhiMiu}
\end{equation}%
whose solution is given in term of bessel functions

\begin{equation}
\phi _{\boldsymbol{\mu }}=\left\{ e^{\frac{5}{2}\beta r\ }\left[ C_{%
\boldsymbol{\mu }}\ J_{5/2}[\frac{\boldsymbol{\mu }}{\beta }e^{\beta r}]+D_{%
\boldsymbol{\mu }}\ Y_{5/2}[\frac{\boldsymbol{\mu }}{\beta }e^{\beta r\ }]%
\right] \right\}
\end{equation}%
That eigenstates are not bounded to the brane and have infinite norm, in
concordance with the result in \cite{Shaposhnikov}.

\section{Maxwell Equations on the Topological Abelian String}

We will now consider the background metric given by (\ref{metric}), where $%
M(r)=L(r)$ is given by (\ref{Mgrande}). Performing variations in the gauge
field, on the action (\ref{EinsteinAH}) the field equations are:

\begin{equation}
-\frac{1}{\sqrt{-G}}\partial _{A}\left[ \sqrt{-G}F^{AB}\right]
=e^{2}\left\Vert \phi \right\Vert ^{2}A^{B}+i\frac{e}{2}(\overline{\phi }%
\partial ^{B}\phi -\phi \partial ^{B}\overline{\phi })  \label{A}
\end{equation}

Here we will assume the Nielsen Olesen anzats (\ref{Amu}) as valid only to
zero order, and look for the equations of the first order fluctuations but
with the simplification given by (\ref{P}). Moreover we will impose axial
symmetry for all initial fields, sources and potential, that means that both
gravity and electromagnetic fields must have axial symmetry. Axial symmetry
jointly with the abelian gauge transformation of the action implies we have
two $U(1)$, one for the gauge field and another for the invariance under
spatial rotations in the 2 dimensional space with coordinates $(r,\theta )$
as in \cite{Giovannini1}\cite{Giovannini2}.

The gauge invariance allows us to impose two gauge fixings%
\begin{eqnarray}
A_{r} &=&0  \label{Ar fixing} \\
A_{\theta } &=&\frac{n}{e}  \label{Atheta fixing}
\end{eqnarray}%
as exact equations at all orders in perturbative theory. Equation (\ref{Ar
fixing}) is equivalent to (\ref{P}). The 4 dimensional Maxwell field is
considered to be a fluctuation from its null background value (\ref{Amu})
\begin{eqnarray*}
A_{\mu } &=&0+\mathcal{A}_{\mu }(x^{\nu },r,\theta ) \\
\mathcal{F}_{\mu \nu } &=&0+\partial _{\mu }\mathcal{A}_{\nu }-\partial _{%
\mathcal{\nu }}\mathcal{A}_{\mu }
\end{eqnarray*}

We will also assume $\phi $ to be a background soliton, so we will look for
solutions that preserve the background condition (\ref{F}):%
\begin{equation}
\phi =v\ f(r)\ e^{in\theta }  \label{FiV}
\end{equation}%
at least at first order. So the scalar field "current" contribution to the
right term in (\ref{A}) vanishes.

Equation (\ref{A}) will be exact to first order in perturbative analysis for
$\mathcal{A}_{\mu }$ with the metric given by the background (\ref{metric})
and (\ref{Mgrande}). Note that in this approximation we have neglected
graviphotons and graviscalars (spin 1 and 0 from fluctuations of the metric)
inspired by the results in \cite{Giovannini1} and \cite{Giovannini2} because
graviphotons and graviscalars were found to be non normalizable either by
their behavior at $r\rightarrow 0$ \ or at $r\rightarrow \infty $. The point
of view here, is that the contribution of the non localized field to the
electromagnetic fluctuation will be weak. As the norm of these wavefunctions
divergesm, the value of the "normalized" wavefunction on the 3 brane and
therefore its superposition with localized fields will be small. Although a
renormalization procedure would exist, graviphoton and graviscalar are first
order fluctuation of the metric (\ref{metric}), and as $\mathcal{F}^{\mu
\theta }$ is itself a first order, then the graviscalar and graviphotons
interaction terms in (\ref{A}) are second order terms and will not be
considered here.

The equation (\ref{A}) for $B=\theta $, due to (\ref{Ar fixing}) and (\ref{F}%
) up to first order is:

\begin{equation*}
-\frac{1}{\sqrt{-G}}\partial _{\mu }\left[ \sqrt{-G}\mathcal{F}^{\mu \theta }%
\right] =\partial _{\mu }\mathcal{F}^{\mu \theta }=0\qquad
\Longleftrightarrow \qquad \partial ^{\theta }\partial _{\mu }\mathcal{A}%
^{\mu }=0
\end{equation*}%
while the equation for $B=r$, due to (\ref{Atheta fixing}) and (\ref{F}) up
to first order is: \ \

\begin{equation*}
\partial _{\mu }\mathcal{F}^{\mu r}=0\qquad \Longleftrightarrow \qquad
\partial ^{r}\partial _{\mu }\mathcal{A}^{\mu }=0
\end{equation*}%
so both equations\ reduces to:%
\begin{equation}
\partial ^{\mu }\mathcal{A}_{\mu }=0  \label{Lorenz}
\end{equation}%
That is the 4 dimensional Lorenz gauge condition and comes from the fact
that the electromagnetic U(1) gauge invariance was broken by the gauge
fixings (\ref{Ar fixing},\ref{Atheta fixing}).

If we add to the action (\ref{EinsteinAH}) and interaction term between an
external current coupled with the gauge field

\begin{equation}
S_{int}=\int dx^{6}\sqrt{-G}\left[ A_{B}J_{ext}^{B}\right]
\label{Interaction}
\end{equation}%
where this "external current" is given by 4 dimensional term "on shell" on
the 3 brane:%
\begin{equation}
J_{ext}^{A}=J_{4D}^{\mu }(x^{\mu },r,\theta )\ \delta _{\mu }^{A}
\label{Jota}
\end{equation}%
only the for the case $B=\mu $ equation (\ref{A}) will acquires a current
term%
\begin{equation}
\partial _{\mu }\mathcal{F}^{\mu \nu }+\frac{1}{\sqrt{-G}}\partial _{r}\left[
\sqrt{-G}\mathcal{F}^{r\nu }\right] +\partial _{\theta }\mathcal{F}^{\theta
\nu }=J_{ext}^{\nu }+v^{2}e^{2}\left\Vert f\right\Vert ^{2}\mathcal{A}^{\nu }
\label{split}
\end{equation}%
that could be break into two equations

\begin{eqnarray}
\partial _{\mu }\mathcal{F}^{\mu \nu } &=&J_{4D}^{\mu }  \label{Maxwell} \\
\frac{-1}{M^{2}L}\partial _{r}\left[ M^{2}L\ \partial _{r}\mathcal{A}_{\mu }%
\right] &=&\frac{1}{R_{o}^{2}L^{2}}\partial _{\theta }^{2}\mathcal{A}_{\mu
}+v^{2}e^{2}\left\Vert f\right\Vert ^{2}\mathcal{A}_{\mu }
\label{wavefunction}
\end{eqnarray}%
So we almost recover the 4 dimensional Maxwell equations with a 4
dimensional source on the brane, in the Lorenz gauge (\ref{Lorenz}), if
equation (\ref{wavefunction}) is accomplished. As the source in equation (%
\ref{Maxwell}) may depends on $(r,\theta )$ we are not exactly recovering
Maxwell, unless the 4 dimensional current $J_{4D}^{\mu }=J_{4D}^{\mu
}(x^{\mu })$ depends only on $x^{\mu }$ or both $\mathcal{F}^{\mu \nu }$ and
$J_{4D}^{\mu }$ has identical warping factors. Then we will recover
completely the 4 dimensional Maxwell equations when $r$ is close to zero. In
this case the photon is simply the 4 dimensional vector potential multiplied
by a warping factor (\ref{acero}) that confines the photon to the 3 brane
universe. To see that the former is indeed the case,we must proceed to the
Fourier analysis of the equations in the following section.

\section{Localizing the Photon Zero Mode on the Topological Abelian String.}

If we expand the gauge field equations in a Fourier series as was done in a
previous section and in \cite{Shaposhnikov} for the tensor case:

\begin{equation}
\mathcal{A}_{\mu }(x^{\nu },r,\theta )=\mathbf{A}_{\mu }(x)\sum_{l}a_{l}(r)\
e^{il\theta }  \label{Fourier}
\end{equation}%
then the equations for the Fourier coefficients using (\ref{Fourier}) in (%
\ref{wavefunction}) is:

\begin{eqnarray}
-\frac{1}{M^{2}L}\frac{\partial }{\partial r}\left[ M^{2}L\ \frac{\partial }{%
\partial r}(a_{l})\right] &=&q_{l}^{2}a_{l}  \label{aele} \\
\text{with \ \ \ \ \ \ }q_{l}^{2}(r) &=&v^{2}e^{2}\left\Vert f(r)\right\Vert
^{2}-\frac{l^{2}}{R_{o}^{2}L(r)^{2}}  \label{qele}
\end{eqnarray}%
\ that is very similar to the equation (\ref{wave}) for the gravity case.

Two important differences arise from equations (\ref{aele}) and (\ref{wave}%
): the first is that in (\ref{wave}) the square mass of the graviton $\mu
^{2}=m_{o}^{2}-\left( \frac{M}{L}\right) ^{2}\frac{\kappa ^{2}}{R_{o}^{2}}$
is a constant eigenvalue due to $M(r)=L(r)$ as in \cite{Shaposhnikov}, while
for (\ref{aele}) the "charge square" (\ref{qele}) is a complicate function.
The second difference is that unlike the graviton case for which $\mu
^{2}\geqslant 0$, in equation (\ref{aele}) there is not physical reason to
avoid $q_{l}^{2}\leq 0.$ In fact as $\lim\limits_{r\rightarrow \infty
}L(r)=0 $ implies that asymptotically $q_{l}^{2}\rightarrow -\infty $ for
all non zero modes $l\neq 0$.

Equation (\ref{aele}) could be written for the massless case for all fourier
modes as

\begin{equation}
a_{l}^{\prime \prime }+3\frac{M^{\prime }}{M}\ a_{l}^{\prime }-q_{l}^{2}\
a_{l}=0  \label{as}
\end{equation}%
here we will take the values for $M(r)$ from equation (\ref{Mgrande}) and $%
f(r)$ from (\ref{ffo}) for the topological abelian string solution obtained
in a previous section and in \cite{GRG2010} :
\begin{equation}
a_{l}^{\prime \prime }-3\beta \tanh \left( \beta r/\delta \right)
a_{l}^{\prime }-\left( v^{2}e^{2}\left\Vert f_{0}\arctan \left( \sinh \
\beta r/\delta \right) \right\Vert ^{2}-\frac{l^{2}/R_{o}^{2}}{\cosh
^{-2\delta }(\beta r/\delta )}\right) a_{l}=0.  \label{numeric}
\end{equation}

Equation (\ref{numeric}) is rather involved even for the zero mode $l=0$
case, so we will use the same approximation (\ref{aproximation}) that was
used in \cite{Shaposhnikov} to calculate the massive modes for the
gravitational case, jointly with the approximation (see (\ref{ffo}))%
\begin{equation}
f(r)\cong \pi \sqrt{\delta }\qquad  \label{faproximation}
\end{equation}%
used in \cite{GRG2010}. Both approximations are\ valid for $r>\frac{\delta }{%
\beta }$, where $\delta $ is a not dimensional parameter related to the
string thickness and $\beta $ the Randall Sundrum warp factor.

\begin{figure}[h]
\begin{minipage}[h]{0.50\linewidth}
\includegraphics[width=7cm]{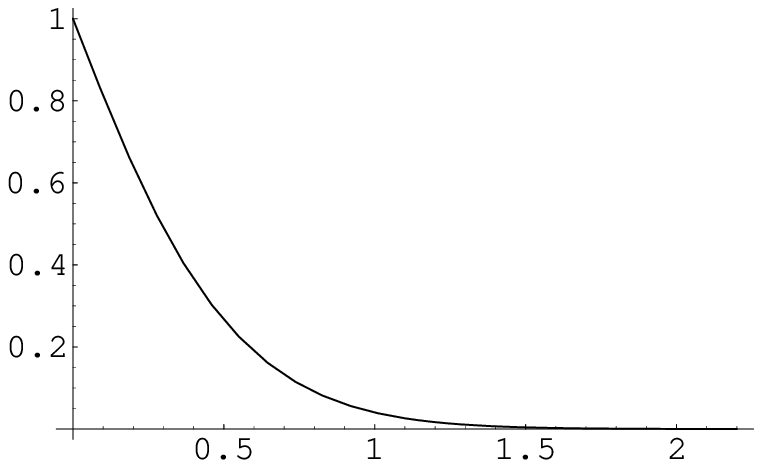}
\end{minipage}%
\begin{minipage}[h]{0.50\linewidth}
\includegraphics[width=7cm]{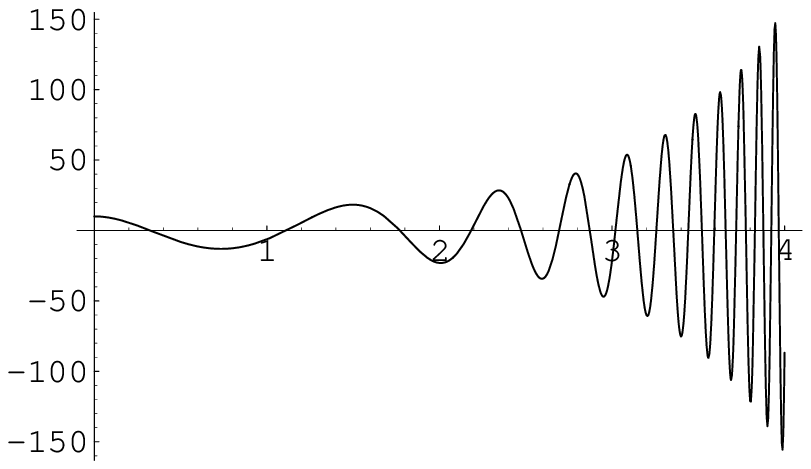}
\end{minipage}
\caption{Numerical integration of eq.(\protect\ref{numeric}) for the
massless modes: $l=0$ (left) and $l=1$ (right) $e=0.1,\ d=50,$ $\protect%
\beta =4\protect\pi /3,\ f_{o}=1.$}
\label{Fig3}
\end{figure}

First we will address the massless zero mode $l=0$ case, later in this
section we will study the massless $l\neq 0$ case. The massive case will be
consider in the next section. Exact numerical solutions could be obtained as
show Fig (\ref{Fig3}).

For the $l=0$ massless case we have

\begin{equation}
a_{0}^{\prime \prime }+3\beta \ a_{0}^{\prime }-q_{0}^{2}\ a_{0}=0,\qquad
\qquad q_{0}^{2}=v^{2}e^{2}\pi ^{2}\delta ,
\end{equation}%
that has the following base of solutions:%
\begin{equation}
a_{0}(r)\propto \exp \left[ \left( \tfrac{3\beta }{2}\pm \sqrt{\left( \tfrac{%
3\beta }{2}\right) ^{2}-q_{0}^{2}}\right) r\right]  \label{acero}
\end{equation}%
with the plus sign (\ref{acero}) is diverging, with the minus sign is
converging provided

\begin{equation}
\delta <\frac{5}{4\pi }\frac{\beta ^{2}}{e^{2}}  \label{ThinLimit}
\end{equation}%
it is also a non oscillating when $\delta <\frac{9}{4\pi }\frac{\beta ^{2}}{%
v^{2}e^{2}}$, that is fulfilled by the former equation (\ref{ThinLimit})
that will be referred as the thin string limit. Note that in the thinnest
limit, when $\delta \rightarrow 0$ the zero mode converges to a constant
value%
\begin{equation*}
\lim\limits_{\delta \rightarrow 0}\;a_{0}(r)\rightarrow 1
\end{equation*}%
but the coefficient can be normalized in the curved space we are working on.

Equation (\ref{aele}) could be rewritten as a Sturm Liouville equation:%
\begin{equation}
\frac{\partial }{\partial r}\left[ p(r)\ \frac{\partial }{\partial r}(a_{l})%
\right] +q(r)\ a_{l}=\rho (r)\ \lambda \ a_{l}  \label{s-l}
\end{equation}%
\begin{equation}
\ p(r)=M^{2}L,\qquad q(r)=M^{2}L\ v^{2}e^{2}\left\Vert f(r)\right\Vert
^{2},\qquad \text{and}\qquad \rho (r)=M(r)  \label{funciones S-L}
\end{equation}%
where the eigenvalues $\ \lambda =l^{2}/R_{o}^{2}$ are positive and real
constants. The operator is self adjoint and the Sturm Lioville problem has
solution with boundary conditions

\begin{equation}
a_{l}^{\prime }(0)=0\qquad \qquad a_{l}^{\prime }(\infty )=0  \label{B.C.1}
\end{equation}%
that are of the same kind of condition taken in \cite{Shaposhnikov}. Note
that the normalization factor comes from $\rho (r)=M(r)$, the weight
function, so the ortonormalization condition is:%
\begin{eqnarray}
\int_{0}^{\infty }dr\ M(r)\ a_{l}(r)\ a_{m}(r) &=&\left\Vert
N_{l}\right\Vert ^{2}\ \delta _{lm}  \notag \\
\left\Vert N_{l}\right\Vert ^{2} &=&\int_{0}^{\infty }dr\ M(r)\ a_{l}^{2}(r)
\label{normaL1}
\end{eqnarray}%
Then the modes could be written in term of a equivalent normalized function
in flat space

\begin{equation}
\mathbf{a}_{\mathbf{l}}^{\mathbf{N}}=\frac{a_{l}(r)}{\left\Vert
N_{l}\right\Vert }\sqrt{M(r)}  \label{anormalized}
\end{equation}%
That is consistent with the normalization used in \cite{Shaposhnikov} and
\cite{GRG2010} for the graviton case, but in graviton case the weight
function was $M^{2}L,$ instead of $M(r)$.

Maxwell equation (\ref{Maxwell}) will be solved in term of the zero mode
only, if we assume that $J_{4D}^{\mu }(x^{\upsilon },r)$ has not dependence
on the angular variable $\theta $
\begin{equation}
\mathcal{A}_{\mu }(x^{\nu },r,\theta )=\mathbf{A}_{\mu }(x)\ \mathbf{a}_{o}^{%
\mathbf{N}}(r)  \label{A0}
\end{equation}%
where

\begin{equation}
\mathbf{a}_{\mathbf{o}}^{\mathbf{N}}\mathbf{(r)}=\frac{a_{0}}{\left\Vert
N_{0}\right\Vert }\cosh ^{-\delta /2}(\beta r/\delta ),\qquad \qquad
\left\Vert N_{0}\right\Vert ^{2}=\cosh ^{-\delta }(\beta r/\delta
)\left\Vert a_{0}\right\Vert ^{2},  \label{Normalization}
\end{equation}%
here $a_{0}$ is given by (\ref{acero}) in the thin limit approximation (\ref%
{ThinLimit}) or could be solved numerically from (\ref{numeric}).

In the null thickness limit $\delta \rightarrow 0$ we simply have

\begin{equation}
\lim_{\delta \rightarrow 0}\mathbf{a}_{\mathbf{o}}^{\mathbf{N}}\rightarrow
\sqrt{\beta }\exp [-\frac{\beta }{2}r].  \label{modozero}
\end{equation}

As the massless zero mode has a $M^{1/2}(r)$ warping factor, to obtain the
equations (\ref{Maxwell}), we must introduce the same warping factor to the
current (\ref{Jota}):
\begin{equation}
J_{ext}^{A}(x^{\mu },r)=J_{4D}^{\mu }(x^{\mu })\ M^{1/2}(r)\ \delta _{\mu
}^{A}  \label{Jota0}
\end{equation}%
So, using (\ref{A0}) and (\ref{Jota0}) in (\ref{Maxwell}) we recover
completely the 4 dimensional Maxwell equations very close to the 3 brane. In
this case the photon is simply the 4 dimensional vector potential multiplied
by a warping factor (\ref{acero}) that confines the photon to the 3 brane
universe.

To obtain the non zero Fourier modes, $l\neq 0$, we could write equation (%
\ref{as}) using (\ref{aproximation}) and (\ref{faproximation}) as:%
\begin{equation}
a_{l}^{\prime \prime }-3\beta \ a_{l}^{\prime }+q_{0}^{2}\ a_{l}=(2^{2\delta
}e^{2\beta r})\ \frac{l^{2}}{R_{o}^{2}}\ a_{l}  \label{adivergente}
\end{equation}%
these equations has the following base of solutions in the thin limit
approximation (\ref{ThinLimit}):%
\begin{equation}
a_{l}(r)\propto \left\{ e^{\frac{3}{2}\beta r\ }I\left[ \pm \tfrac{1}{\beta }%
\sqrt{\left( \tfrac{3\beta }{2}\right) ^{2}-q_{0}^{2}},\tfrac{2^{\delta }l}{%
\beta R_{o}^{2}}e^{\beta r\ }\right] \right\}  \label{Bessel-I}
\end{equation}%
where the notation $I[n,x]=I_{n}[x]$ has been used for Bessel functions.

These functions are not bounded for $r\rightarrow \infty $, and could not be
normalized because integrals in (\ref{normaL1}) are diverging. They neither
can be normalized using the graviton norm $M^{\frac{3}{2}}(r)$ \ as in (\ref%
{normaG}). So massless non zero modes are not localized on the world brane
and wanders into the bulk. The massless zero mode is bounded to the 3-brane
only if (\ref{ThinLimit}) is accomplished. Finally we want to stand out that
if (\ref{ThinLimit}) is not accomplished, $a_{0}$ will be an exponential
growing factor and the photon wavefunction will not longer be localized on
the world brane, and scape into the bulk.

\section{Proca Massive Modes and the Corrections to the Coulomb Potential}

The orthogonality of Fourier modes implies that equation (\ref{split}), and
therefore (\ref{Maxwell}) and (\ref{wavefunction}) must be solved order by
order. The complete solution will be a superposition of the eigenfunctions
with eigenvalues $\mu $ and Fourier index $l$%
\begin{equation}
\mathcal{A}_{\mu }^{(l)}(x^{\nu },r,\theta )=\mathbf{A}_{\mu }^{(l)}(x)\
a_{l}(r)\ e^{il\theta },  \label{eigenfunction}
\end{equation}%
that is slightly more general that (\ref{Fourier}), and could be understood
as a collection of vector potentials, one for each mode, with
electromagnetic fields%
\begin{equation}
\mathcal{F}_{\mu \nu }^{(l)}(x^{\nu },r,\theta )=\mathbf{F}_{\mu \nu
}^{(l)}(x)\ a_{l}(r)\ e^{il\theta }.
\end{equation}

Then equation (\ref{split}) with not current term ($J_{ext}^{\nu }=0$) could
be written as
\begin{equation}
\frac{a_{l}(r)}{M^{2}}\eta ^{\nu \upsilon }\partial _{\upsilon }\mathbf{F}%
_{\nu \mu }^{(l)}-\frac{1}{M^{2}L}\frac{\partial }{\partial r}\left[ M^{2}L\
\frac{\partial }{\partial r}[a_{l}(r)]\right] \mathbf{A}_{\mu }^{(l)}(x)=%
\left[ v^{2}e^{2}\left\Vert f(r)\right\Vert ^{2}-\frac{l^{2}}{R_{o}^{2}\
L^{2}(r)}\right] a_{l}(r)\mathbf{A}_{\mu }^{(l)}(x)
\end{equation}%
the sign in $-\frac{l^{2}}{R_{o}^{2}L^{2}}$ is what leads to a non
normalizable expressions as (\ref{Bessel-I}).

Note that we can add to both right and left sides a Proca mass term:

\begin{equation}
\frac{a_{l}(r)}{M^{2}}\left( \square \mathbf{A}_{\mu }^{(l)}+m^{2}\mathbf{A}%
_{\mu }^{(l)}\right) -\frac{1}{M^{2}L}\frac{\partial }{\partial r}\left[
M^{2}L\ \frac{\partial }{\partial r}[a_{l}(r)]\right] \mathbf{A}_{\mu
}^{(l)}(x)=\left[ q_{0}^{2}+\frac{m^{2}-l^{2}/R_{o}^{2}}{L^{2}}\right]
a_{l}(r)\mathbf{A}_{\mu }^{(l)}(x)  \label{splitmasivo}
\end{equation}%
where (\ref{faproximation}) and Lorenz gauge (\ref{Lorenz}) have been used.

Former equation (\ref{splitmasivo}) could be split in similar way to (\ref%
{Maxwell}) and (\ref{wavefunction}) as%
\begin{eqnarray}
\square \mathbf{A}_{\mu }^{(l)}+m^{2}\mathbf{A}_{\mu }^{(l)} &=&0,
\label{proca} \\
-\frac{1}{M^{2}L}\frac{\partial }{\partial r}\left[ M^{2}L\ \frac{\partial }{%
\partial r}[a_{l}(r)]\right] &=&\left[ q_{0}^{2}+\frac{m^{2}-l^{2}/R_{o}^{2}%
}{L^{2}}\right] a_{l}(r).  \label{regularized}
\end{eqnarray}%
Equation (\ref{proca}) is a Proca equation with 4 dimensional mass $m$ for
the photon. Equation (\ref{regularized}) gives a regularized version of (\ref%
{aele}) when $m^{2}-l^{2}/R_{o}^{2}>0$ and is a self adjoint Sturm Liouville
equation.

We could write equation (\ref{aele}) using \ref{Mgrande} and approximations (%
\ref{aproximation}):

\begin{equation}
a_{l}^{\prime \prime }-3\beta \ a_{l}^{\prime }+q_{0}^{2}\ a_{l}+(2^{2\delta
}e^{2\beta r})(m^{2}-\ \frac{l^{2}}{R_{o}^{2}})\ a_{l}=0
\label{aconvergente}
\end{equation}%
these equations has the following base of solutions:%
\begin{equation}
a_{l}(r)\propto e^{\frac{3}{2}\beta r\ }\left\{ J\left[ \tfrac{1}{\beta }%
\sqrt{\left( \tfrac{3\beta }{2}\right) ^{2}-q_{0}^{2}},\frac{2^{\delta }%
\boldsymbol{\mu }}{\beta }e^{\beta r\ }\right] ,Y\left[ \tfrac{1}{\beta }%
\sqrt{\left( \tfrac{3\beta }{2}\right) ^{2}-q_{0}^{2}},\frac{2^{\delta }%
\boldsymbol{\mu }}{\beta }e^{\beta r\ }\right] \right\}  \label{Bessel-.J}
\end{equation}%
where the eigenvalues $\boldsymbol{\mu \ }\mathbf{=}\sqrt{m^{2}-\ \frac{l^{2}%
}{R_{o}^{2}}}>0$ are effective mass term and notation $J[n,x]=J_{n}[x]$ and $%
Y[n,x]=Y_{n}[x]$ \ has been used for Bessel functions

To obtain the first order electromagnetic field complete solution, we must
sum over all the Fourier modes $l$, and as (\ref{proca}) and (\ref%
{regularized}) depends on the mass, we must sum also sum over all the masses
$m>l/R_{o}^{2}$, or equivalently on the eigenvalues $\boldsymbol{\mu }$ as
indicated:

\begin{equation}
\mathcal{A}_{\sigma }(x^{\nu },r,\theta )=\sum_{l}\sum\limits_{{\small m>l/R}%
_{{\small o}}^{{\small 2}}}\mathbf{A}_{\sigma }^{(l)}(x)\ a_{l}(r)\
e^{il\theta }=\sum_{l}\sum\limits_{\boldsymbol{\mu }>0}\mathbf{A}_{\sigma
}^{(l)}(x)\ a_{\boldsymbol{\mu }}(r)\ e^{il\theta }  \label{sumatoria}
\end{equation}

The expression for (\ref{Bessel-.J}) could be simplified in null string
thickness $\delta \rightarrow 0$ to:

\begin{equation}
a_{\boldsymbol{\mu }}(r)=e^{\frac{3}{2}\beta r\ }\left[ C_{\boldsymbol{\mu }%
}\ J_{3/2}[\frac{\boldsymbol{\mu }}{\beta }e^{\beta r}]+D_{\boldsymbol{\mu }%
}\ Y_{3/2}[\frac{\boldsymbol{\mu }}{\beta }e^{\beta r\ }]\right]
\label{ajota}
\end{equation}%
where a more standard notation for Bessel functions has been introduced and $%
C_{\boldsymbol{\mu }}$ and $D_{\boldsymbol{\mu }}$ are constants. That
expression is quite similar to that for gravitons \cite{Shaposhnikov} and
domain walls \cite{RS1}\cite{RS2}\cite{callinRanvdal} changing only the type
of the Bessel function and the exponent in the warp factor.

In the limit $r\rightarrow \infty $, the solution (\ref{ajota}) grow
exponentially for non zero $\boldsymbol{\mu }$. The standard way to
regularize this, is to introduce a finite radial distance cutoff $\ r_{\max
} $ where the boundary condition (\ref{B.C.1}) will be imposed for $%
r=r_{\max } $ instead of $r=\infty $

\begin{equation}
a_{\boldsymbol{\mu }}^{\prime }(0)=0\qquad \qquad a_{\boldsymbol{\mu }%
}^{\prime }(r_{\max })=0  \label{B.C.2}
\end{equation}%
These boundary conditions, jointly with the normalization condition (\ref%
{normaL1}) leads to a discrete spectrum

\begin{equation}
\boldsymbol{\mu }_{\mathbf{n}}\approx \beta (\mathbf{n}+1)\pi \ e^{-\beta
r_{\max }\ }  \label{spectrum}
\end{equation}%
for enough large $\mathbf{n\in
\mathbb{Z}
,}$ and solving for the constants we get:%
\begin{eqnarray}
C_{\boldsymbol{\mu }_{\mathbf{n}}} &=&\frac{1}{N_{\mathbf{n}}}Y_{1/2}[\frac{%
\boldsymbol{\mu }_{\mathbf{n}}}{\beta }]  \notag \\
D_{\boldsymbol{\mu }_{\mathbf{n}}} &=&\frac{-1}{N_{\mathbf{n}}}J_{1/2}[\frac{%
\boldsymbol{\mu }_{\mathbf{n}}}{\beta }]  \label{C2}
\end{eqnarray}

When the cutoff is imposed the sum in Fourier expansion (\ref{sumatoria})
will be replaced by a sum over $\mathbf{n,}$\ because (\ref{spectrum}) are
the only acceptable values of $\boldsymbol{\mu }$\textbf{\ }that will lead
to renormalizable wave functions as expected by (\ref{normaL1}).

The normalization (\ref{normaL1}) of the massive modes, can be calculated
using (\ref{ajota}),(\ref{C2}), as:
\begin{align}
\left\Vert N_{\mathbf{n}}\right\Vert ^{2}& =\int\limits_{0}^{r_{\max }}dr\
e^{2\beta r\ }(Y_{1/2}[\frac{\boldsymbol{\mu }_{\mathbf{n}}}{\beta }]\
J_{3/2}[\frac{\boldsymbol{\mu }_{\mathbf{n}}}{\beta }e^{\beta r\ }]-J_{1/2}[%
\frac{\boldsymbol{\mu }_{\mathbf{n}}}{\beta }]\ Y_{1/2}[\frac{\boldsymbol{%
\mu }_{\mathbf{n}}}{\beta }e^{\beta r\ }])^{2}  \notag \\
& =\frac{(J_{1/2}[\frac{\boldsymbol{\mu }_{\mathbf{n}}}{\beta }%
])^{2}+(Y_{1/2}[\frac{\boldsymbol{\mu }_{\mathbf{n}}}{\beta }])^{2}}{\pi
\boldsymbol{\mu }_{\mathbf{n}}}(e^{\beta r_{\max }\ }-1)  \label{normaN2}
\end{align}%
where the asymptotical form for $J_{3/2}[\frac{\boldsymbol{\mu }_{\mathbf{n}}%
}{\beta }e^{\beta r}]$ and $Y_{3/2}[\frac{\boldsymbol{\mu }_{\mathbf{n}}}{%
\beta }e^{\beta r}]$, was used due to the large factor $e^{\beta r}$ for
large $r\rightarrow r_{\max }$ and enough large $\mathbf{n}$

When the cutoff is imposed, the sum in (\ref{Fourier}) and (\ref{sumatoria})
turns out to be over $\mathbf{n\in
\mathbb{Z}
}$. So it will be used from now on in the rest of this work. Therefore,
massive electromagnetic modes have fourier coefficients given by (\ref%
{anormalized}):

\begin{equation}
\mathbf{a}_{\mathbf{n}}^{N}(r)=\frac{e^{\frac{3}{2}\beta r\ }}{\left\Vert N_{%
\mathbf{n}}\right\Vert }\left[ Y_{1/2}[\frac{\boldsymbol{\mu }_{\mathbf{n}}}{%
\beta }]\ J_{3/2}[\frac{\boldsymbol{\mu }_{\mathbf{n}}}{\beta }e^{\beta
r}]-J_{1/2}[\frac{\boldsymbol{\mu }_{\mathbf{n}}}{\beta }]\ Y_{3/2}[\frac{%
\boldsymbol{\mu }_{\mathbf{n}}}{\beta }e^{\beta r\ }]\right]
\label{amassive}
\end{equation}

A Schr\"{o}dinger equation could be obtained for the massive modes making
the change:

\begin{equation}
u_{\mathbf{n}}(r)=\mathbf{a}_{\mathbf{n}}^{N}(r)\ M^{-1}  \label{uN}
\end{equation}%
that accomplishes%
\begin{eqnarray}
-u_{\mathbf{n}}^{^{\prime \prime }}(r)+V_{EM}.u_{\mathbf{n}}(r) &=&0
\label{schrodinger} \\
V_{EM} &=&\frac{3}{2}\frac{M^{\prime \prime }}{M}+\frac{3}{4}\left( \frac{%
M^{\prime }}{M}\right) ^{2}-q_{l}^{2}(r)  \notag
\end{eqnarray}

The potential could be found calculating the integral of the Feynman graph
at the tree level, for the interchange of a virtual photon between two
stationary charges $q_{1}$and $q_{2}$ placed on the 3-brane at $r=0,$ in the
limit in which the photon energy $k^{0}\rightarrow 0$ goes to zero:

\begin{equation}
V(R)=\frac{q_{1}\ q_{2}}{\beta R}\left\{ \left\Vert u_{\mathbf{0}%
}(r=0)\right\Vert _{Schr}^{2}+pf\sum_{l}\sum_{\mathbf{n}>0}\left\Vert u_{%
\mathbf{n}}(r=0)\right\Vert _{Schr}^{2}\exp (-m\ R)\right\}  \label{Coulomb2}
\end{equation}%
where $R=\sqrt{\left\Vert \vec{x}\right\Vert ^{2}}$is the spatial 3-distance
between the charges, $m=\sqrt{\boldsymbol{\mu }_{\mathbf{n}%
}^{2}+l^{2}/R_{o}^{2}}$ is the Proca mass, $\left\Vert u\right\Vert
_{Schr}^{2}$ is the usual Schr\"{o}dinger norm $\int dx^{4}u^{\ast }u$ and $%
pf$ is a polarization factor.

For both massless and Proca photon the polarization factor is:
\begin{equation}
\lim\limits_{k^{0}\rightarrow 0}pf=g^{00}-\frac{k^{0}k^{0}}{\boldsymbol{\mu }%
_{\mathbf{n}}}=1  \label{polarization}
\end{equation}%
where $k_{\mu }$ the momentum of the virtual photon, is the conjugate
Fourier Transform variable to $x^{\mu }$.

Note that (\ref{Coulomb2}) takes into account the contribution of the
Maxwell or zero non massive propagator $\frac{1}{R}$ and the contribution of
the massive or Proca propagator $\frac{\exp (-m\ R)}{R}$ weighted by the
squared norm of its wavefunction (\ref{uN}), that is its probability of been
found at $r=0$. Although the norm of the massive modes (\ref{amassive}) is
infinite, due to the large factor $e^{\beta r_{\max }\ }$ in equation (\ref%
{normaN2}), a finite limit could be found by considering the infinite sum in
(\ref{Coulomb2}).Using that $\lim\limits_{r\rightarrow 0}[u_{\mathbf{n}%
}(r)]= $ $\mathbf{a}_{\mathbf{n}}^{\mathbf{N}}(0),$because $M(0)=1,$
therefore the correction to the Coulomb potential could be calculated as:

\begin{equation}
\mathbf{\triangle }V=\frac{q_{1}\ q_{2}}{R}\left\{ \sum_{l}\sum_{\mathbf{n}%
>0}\frac{\left\Vert \mathbf{a}_{\mathbf{n}}^{\mathbf{N}}(r=0)\right\Vert
_{Schr}^{2}}{\left\Vert \mathbf{a}_{\mathbf{0}}^{\mathbf{N}}(r=0)\right\Vert
_{Schr}^{2}}\exp (-m\ R)\ \mathbf{\triangle n}\right\}  \label{correction}
\end{equation}%
This infinite sum could be transformed into an integral over the masses $%
\boldsymbol{\mu }_{\mathbf{n}}>0$, using the quantization condition (\ref%
{spectrum}) for which:
\begin{equation}
\mathbf{\triangle n}=\frac{1}{\beta \pi }e^{\beta r_{\max }}\mathbf{%
\triangle }\boldsymbol{\mu }_{\mathbf{n}}\rightarrow \frac{1}{\beta \pi }%
e^{\beta r_{\max }}\mathbf{d}\boldsymbol{\mu }  \label{deltaN}
\end{equation}%
So we get to an integral over the masses of the modes:
\begin{equation}
\mathbf{\triangle }V=\frac{q_{1}\ q_{2}}{R}\sum_{l}\int\limits_{\boldsymbol{%
\mu }\mathbf{>0}}\frac{\left\Vert \mathbf{a}_{\mathbf{n}}^{\mathbf{N}%
}(r=0)\right\Vert _{Schr}^{2}}{\left\Vert \mathbf{a}_{\mathbf{0}}^{\mathbf{N}%
}(r=0)\right\Vert _{Schr}^{2}}\exp (-\sqrt{\boldsymbol{\mu }%
^{2}+l^{2}/R_{o}^{2}}\ R)\ \frac{e^{\beta r_{\max }}}{\beta \pi }\mathbf{d}%
\boldsymbol{\mu }  \label{correction2}
\end{equation}

The former integral could be obtained using (\ref{normaN2}), (\ref{amassive}%
), (\ref{uN}) and (\ref{modozero}). Taking either the large or small limit
for $\mathbf{\mu }_{\mathbf{n}}$ in (\ref{normaN2}) we get:

\begin{equation}
\left\Vert N_{\mathbf{n}}\right\Vert ^{2}\cong \frac{2\beta \ e^{\beta
r_{\max }}}{\pi ^{2}\boldsymbol{\mu }_{\mathbf{n}}^{2}}  \label{Nm}
\end{equation}%
Evaluating (\ref{amassive}) at the brane, $r=0,$ and using (\ref{Nm}) and
Bessel identities we simply obtain:%
\begin{equation*}
\left\Vert \mathbf{a}_{\mathbf{n}}^{\mathbf{N}}(0)\right\Vert ^{2}\cong
\frac{2\beta }{e^{\beta r_{\max }}}
\end{equation*}%
and using the thin limit for the zero mode (\ref{modozero}) $\mathbf{a}_{%
\mathbf{0}}^{\mathbf{N}}(0)=\sqrt{\beta },$ we obtain:

\begin{equation}
\mathbf{\triangle }V=\frac{q_{1}\ q_{2}}{R}\sum_{l}\int\limits_{\mathbf{0}%
}^{\infty }\mathbf{d}\boldsymbol{\mu }\ \frac{2}{\beta \pi }\exp (-\sqrt{%
\boldsymbol{\mu }^{2}+l^{2}/R_{o}^{2}}\ R)  \label{correction3}
\end{equation}%
Note that in the former integral the cutoff factor $e^{\beta r_{\max }}$ was
simplified, so the result is cutoff independent. Instead of equation (\ref%
{modozero}), equations (\ref{acero}) and (\ref{Normalization}) could be
used, making a Taylor expansion in $\delta $ gives the same result at second
order term in perturbative theory. In the sum there is implicit a step
factor $\mathbf{\triangle }l=1,$ if we divide and multiply by $R_{o}$ we get:

\begin{equation}
\mathbf{\triangle }V=\frac{q_{1}\ q_{2}}{R}\left[ \lim_{\Upsilon \rightarrow
\infty }\sum_{l=0}^{\Upsilon }\frac{\mathbf{\triangle }l}{R_{o}}\int\limits_{%
\mathbf{0}}^{\infty }\mathbf{d}\boldsymbol{\mu }\ \exp (-\sqrt{\boldsymbol{%
\mu }^{2}+l^{2}/R_{o}^{2}}\ R)\right] \frac{2R_{o}}{\beta \pi }
\label{correction4}
\end{equation}%
That in the limit $R_{o}>>\mathbf{\triangle }l$ and making the change of
variables%
\begin{eqnarray*}
\frac{l}{R_{o}} &=&\mathbf{Y,}\qquad \qquad \frac{\mathbf{\triangle }l}{R_{o}%
}\rightarrow \mathbf{dY,} \\
\boldsymbol{\mu }\mathbf{=X,} &&\qquad \qquad \mathbf{d}\boldsymbol{\mu }=%
\mathbf{dY}
\end{eqnarray*}%
transform (\ref{correction4}) into a double integral%
\begin{equation}
\mathbf{\triangle }V=\frac{q_{1}\ q_{2}}{R}\left[ \frac{2R_{o}}{\beta \pi }%
\int\limits_{\mathbf{0}}^{\infty }\int\limits_{\mathbf{0}}^{\infty }\mathbf{%
dX\ dY}\ \exp (-\sqrt{\mathbf{X}^{2}+\mathbf{Y}^{2}}\ R)\right]
\label{correction5}
\end{equation}%
that could be calculated changing to polar coordinates with radius $%
\boldsymbol{\rho }=\sqrt{\mathbf{X}^{2}+\mathbf{Y}^{2}}$ and integrating
over the area of the upper left quarter of the plane between $[0,\pi /2]$.
So finally:%
\begin{equation}
\mathbf{\triangle }V=\frac{q_{1}\ q_{2}}{R}\left[ \frac{R_{o}}{\beta }%
\int\limits_{\mathbf{0}}^{\infty }\mathbf{\ }\boldsymbol{d\rho }\ \exp (-%
\boldsymbol{\rho }\mathbf{\ }R)\right] =\frac{q_{1}\ q_{2}}{R}\left[ \frac{%
R_{o}}{\beta \ R^{2}}\right]  \label{final}
\end{equation}%
If we have a non warped, flat space with 5 dimensional spatial coordinates,
a $\frac{1}{R^{3}}$ would be expected for the Coulomb potential using Gauss
law that is coincident with our result.

\section{Summary and outlook}

In this work, the confinement of electromagnetic field is studied in axial
symmetrical, 6D warped World Brane, using recently proposed \cite{GRG2010}
topological abelian string vortex solutions as background. The field
equations were calculated only to first order in perturbative analysis for the vector
field. The metric field was assumed to be exact and the solitonic scalar solution was assumed 
as solitonic background. As the theory is assumed to be
axial symmetric and also containing classical electrodynamic or Maxwell
theory, we have a $U(1)\times U(1)$ invariant theory as in \cite{Giovannini1}
and \cite{Giovannini2}, \ that allow us to make two gauge U(1) fixings (\ref%
{Ar fixing}) and (\ref{Atheta fixing}), that is consistent with the
topological abelian string solution, and lead to simplifications in the spin
1 fluctuation equations.

There are several conclusions we found throughout this work:
1. There is a massless, spin 1, Fourier zero mode, bounded to the 3-brane
universe, with a warping factor in the bulk if the string-vortex is thin
enough. The shape of this mode could be seen in Fig(\ref{Fig3}) from
numerical integration, or obtained in the thin string approximation (\ref%
{ThinLimit}) using (\ref{A0}), (\ref{acero}) and (\ref{Normalization}) even could be
simplified \ to (\ref{modozero}) in the null thickness limit $\delta
\rightarrow 0$.

2. The massless zero mode is consistent with Maxwell equations (\ref{Maxwell}%
) with an external current warped throughout the bulk.

3. All other modes in the photon expansion (\ref{eigenfunction}) are\ not
localized and massive. They follow Proca equations (\ref{proca}).

4. The main conclusion of this work is that the correction to the Coulomb
law, produced by the massive modes, in the thin string limit\bigskip\ is:%
\begin{equation}
V(r)\cong \frac{q_{1}\ q_{2}}{R}\left( 1+\frac{R_{o}\ R_{rs}}{R^{2}}\right)
\label{coulomb}
\end{equation}%
The expected Coulomb potential for flat 6D spacetime is $\frac{1}{R^{3}}$,
but in this case, it also depends on the factor $R_{o}$, that could be seen
as the radius of compactification of the angular coordinate, and $R_{rs}=%
\frac{1}{\beta }$ that is the distance at which the metric falls by a factor
of $\exp (-1)$ due to the warping factor in Randall Sundrum theories. The
value of $R_{rs}$ is unknown, and could be in a very wide range, as
short as several Planck length \cite{RS1} up to the experimental limit for
Newton potential around$\ 10^{-2}$cm \cite{submilimeter}. The value of $%
R_{o} $ is also not know, but must be larger than Planck length, and shorter
than the experimental limit of validity for the Coulomb law of$\ 10^{-16}$cm
\cite{Lab}\cite{astronomyMU}. Experiments to test at short distances the
electric Coulomb potential, could be used to establish the existence or not of extra warped
dimensions at distances of $10^{-18}$cm in the near future. That is an
increase by a factor of $10^{15}$ with respect to the actual capacity to
observe gravitational effects due to the warping or compactness of extra
dimensions.

Observable effects of the correction to the Coulomb potential (\ref{coulomb}%
) implies at least a deviation of $\frac{\mathbf{\triangle }V}{V}=0.1$,
because this ensures a new decimal figure that could not be explained by $%
\frac{1}{R}$ potential alone. In the most optimistic scenario: assuming $%
R_{o}\approx 10^{-18}$cm and $\frac{\mathbf{\triangle }V}{V}=0.1$, in order
to obtain an observable change in Coulomb law at a distance of $R\approx
10^{-17}$cm (that could be achieve at the LHC) implies that $R_{rs}$ must be
greater than $10^{-17}$cm. But if $R_{o}\approx 10^{-18}$cm and $%
R_{rs}\approx 10^{-28}$cm, for a distance of $R\approx 10^{-17}$cm then we
obtain that $\frac{\mathbf{\triangle }V}{V}=10^{-12}$ so the precision of
the Coulomb law could be amazing.

Proca photons has a long history in both theoretical and experimental
physics, a striking limit of $10^{-48}$gr for the photon mass \cite%
{astronomyMU} has been set by astronomical measures. As we have seen in the previous 
sections, most of the photons
will correspond to the massless zero mode, that are localized at $r=0$, just
over the 3-brane universe. On the other hand, massive photons are not
bounded to the 3-brane universe and wander in the extra dimensions, throughout the bulk, so the 
probability of catch one of the Proca photons is extremely low. Lab 
experiments \cite{Lab} and astronomical measurements
\cite{astronomyMU} limiting the photon mass, assume that all photons has the
same small mass. As this is not the case for this model, these mass limits
do not apply.


\section*{Acknowledgments}

This work was supported by CDCHT-UCLA under project 020-CT-2009.



\end{document}